\documentclass[10pt, oneside, a4paper]{article}
\usepackage{latexsym}
\usepackage{amssymb}
\usepackage{amsthm}
\usepackage{cite}
\newcommand {\be} {\begin{equation}}
\newcommand {\ee} {\end{equation}}
\newcommand {\bea} {\begin{eqnarray}}
\newcommand {\eea} {\end{eqnarray}}
\newcommand {\bean} {\begin{eqnarray*}}
\newcommand {\eean} {\end{eqnarray*}}
\theoremstyle{plain}

\newcommand {\nn} {\nonumber}
\newcommand {\sss} {\scriptscriptstyle}

\newcommand {\we} {\wedge}
\newcommand {\fs}{\footnotesize}

\newcommand {\ga} {\gamma}

\newcommand {\de} {\delta}

\newcommand {\ep} {\epsilon}

\newcommand {\la} {\lambda}
\newcommand {\La} {\Lambda}
\newcommand {\om} {\omega}

\newcommand {\Si} {\Sigma}

\newcommand {\Up} {\Upsilon}

\newcommand {\gh} {\ensuremath{gh_{\#}}}
\newcommand
{\funcright}[3]{\frac{\stackrel{\leftarrow}{\de}}{\de {#1^{\sss
#2}#3}}}
\newcommand
{\funcleft}[3]{\frac{\stackrel{\rightarrow}{\de}}{\de {#1_{\sss
#2}#3}}}
\newcommand{\rightder}[3]{#1\frac{\stackrel{\leftarrow}{\partial}}{\partial{#2^{\sss
#3}}}}
\newcommand{\leftder}[3]{\frac{\stackrel{\rightarrow}{\partial}}{\partial{#2^{\sss
#3}}}#1}

\newcommand{\bdm}{\begin{displaymath}}
\newcommand{\edm}{\end{displaymath}}

\begin{document}
\begin{titlepage}
\vspace*{40mm}
\begin{center}
 {\LARGE\bf Superfield algorithm for\\}
\end{center}\vspace*{0.0 mm}
\begin{center}
{\LARGE\bf higher order gauge field theories }
\end{center} \vspace*{0.0 mm}
\vspace*{3 mm} \vspace*{3 mm}
\begin{center}Ludde Edgren\footnote{E-mail:
edgren@fy.chalmers.se} and Niclas Sandstr\"om\footnote{E-mail:
tfens@fy.chalmers.se}
 \\ \vspace*{7 mm} {\sl Department of Theoretical Physics\\ Chalmers University of Technology\\
G\"{o}teborg University\\
S-412 96  G\"{o}teborg, Sweden}\end{center} \vspace*{25 mm}
\begin{center}
\begin{abstract}
\vspace*{1 mm}
We propose an algorithm for the construction of higher order gauge
field theories from a superfield formulation within the
Batalin-Vilkovisky formalism. This is a generalization of the superfield algorithm recently
considered by Batalin and Marnelius. This generalization seems to allow
for non-topological gauge field theories as well as alternative
representations of topological ones. A five dimensional
non-abelian Chern-Simons theory and a topological Yang-Mills theory are treated as examples.

\end{abstract}
\end{center}
\end{titlepage}
\noindent The superfield algorithm introduced by Batalin and
Marnelius \cite{Batalin:2001ge, Batalin:2001su} provides a general
algorithm for constructing a class of first order gauge field
theories. The Batalin-Vilkovisky (BV) formalism \cite{Batalin:1981ga, Batalin:1983qu}
is there used as a framework for generating consistent quantum
gauge field theories (at least up to gauge fixing and renormalization).
In line with AKSZ \cite{Alexandrov:1995ge}, topological gauge field theories are naturally
incorporated in the class of theories generated by this superfield
algorithm. For instance, four and six dimensional Schwarz-type
topological field theories which were constructed by means of the superfield
algorithm in \cite{Edgren:2002fi}.

In this communication we propose a generalization of the superfield
algorithm to allow for higher order gauge field theories. This is mainly achieved by
introducing non-dynamical super multiplier fields. This
generalized version may also allow for non-topological theories,
suggested by our construction of a five dimensional Chern-Simons
theory. All higher dimensional ($n\geq 5$) Chern-Simons
theories generically possess local degrees of freedom according to
\cite{Banados:1995lo, Banados:1996dy}. Possible theories are generated from a master action in superspace
which is obtained by a ghost number prescription and a
simple local master equation, which in turn is a generalization of the one considered in
\cite{Batalin:2001ge, Batalin:2001su}. The ordinary fields and
antifields are components of the superfields, such that original
gauge theories are obtained by a straightforward reduction procedure. To obtain a
consistent quantum gauge field theory, a gauge fixing of the
master action is necessary in addition to other issues such as for
example renormalization. These issues are not treated in this
paper.

In the first section we introduce a generalized master action and the
master equation that follows from it. We consider the
boundary conditions that this master action has to satisfy in
order to define a possible consistent gauge field theory. A key ingredient
in the superfield algorithm considered in \cite{Batalin:2001ge,
Batalin:2001su} is the above mentioned local master equation obtained from
the ordinary BV master equation in superfield form. Also here we
consider a local master equation which is a generalization of the one in \cite{Batalin:2001ge,
  Batalin:2001su} and we propose a generalized superfield algorithm
for generating higher order gauge field theories as solutions to this new local
master equation. Section 2 gives a description of the proposed generalized
superfield algorithm and the motivation for its ingredients. In
section 3 we construct a topological Yang-Mills theory by
means of this generalized superfield algorithm. Section 4 and 5
continue by discussing the construction of three and five dimensional
non-abelian Chern-Simons theories.

We use the same notation and conventions as the
ones considered in \cite{Edgren:2002fi}.

\section{General considerations}
Consider a master action $\Si$ written as a field theory living on
a $2n$-dimensional supermanifold $\mathcal M$, with $n$ Grassmann
odd and $n$ Grassmann even dimensions, \be\label{masteraction}
    \Si=\int_{\mathcal M}{d^{\sss n}ud^{\sss{n}}\tau
    \mathcal{L}_n(u,\tau)}.
 \ee
The supermanifold $\mathcal M$ is coordinatized by
$(u^a,\,\tau^a)$, where $a=\{1,...,n\}$ and $u^a$ denotes the
Grassmann even and $\tau^a$ the Grassmann odd coordinates
respectively. The Lagrangian density $\mathcal L_n$ is chosen to have the form
\bea \label{Lagrangedensity}
    \mathcal{L}_n(u,\tau)=K^*_{\sss P}(u,\tau)DK^{\sss P}(u,\tau)(-1)^{\ep_{\sss{P}}\sss{+n}}-S(K^*_{\sss P},K^{\sss
    P}, DK^*_{\sss P},DK^{\sss P}).
 \eea
$K^{\sss P}(u,\tau)$ represents the superfields and
$K^*_{\sss P}(u,\tau)$ the associated superfields, defined by
(\ref{Lagrangedensity}) and the properties below. $S$ is local in the
superfields and $D$ is the de Rham differential, $D^2=0$, with \be\label{de_Rham}
    D:=\tau^a\frac{\partial}{\partial u^a}.
\ee 
The chosen Lagrangian is the same as the one considered in
\cite{Batalin:2001ge, Batalin:2001su}, but here we allow terms
in $S$ involving the derivatives $DK, DK^*$. $K^{\sss P}$ and
$K^*_{\sss P}$ denote the superfields and associated superfields
but they also represent the super multiplier fields $\La^{\sss P},
\La^*_{\sss P}$, which are restricted by $D\La^{\sss P}=0$ and $D\La^*_{\sss P}=0$.
Thus they are non-dynamical fields entering in the interaction term
$S$, but not in the kinetic term $K^*_{\sss P}DK^{\sss P}$. A
motivation for these properties of the multiplier fields is given
in section 3.

The Grassmann parities of the associated superfields are
 \be\label{parity}
    \ep(K^*_{\sss P})=\ep_{\sss P}+n+1 \quad \textrm{where} \quad
    \ep(K^{\sss P}):=\ep_{\sss P}.
 \ee
This follows from the master action $\Si$, since this should be an
even functional. The Grassmann parities of the super multiplier
fields can be identified once the master action $\Si$ has been
specified. The conditions are that $\Si$ have ghost number zero and
the odd coordinate $\tau$ carry ghost number one and this implies ghost
number restrictions on the superfields and corresponding
associated superfields,
 \be\label{ghostnumber}
   \gh K^{\sss P}+\gh K^*_{\sss P}=n-1.
 \ee
It also follows that the local function $S$ carries ghost number and Grassmann parity,
\be
\gh S=n, \qquad \ep(S)=n.
\ee
The cornerstone of the BV-formalism is the
classical master equation, requiring the master action $\Si$ to satisfy,
\be
\label{classicalmaster}
    (\Si,\Si)=0,
 \ee
where the bracket is the standard BV-bracket. Here this master
 equation may be written in a superfield form since the standard
 BV-bracket may be defined by
\bea \label{funcbracket}
    ({F},{G})&:=&
    \int_{\mathcal M}{{F}\funcright{K}{P}{(u,\tau)}(-1)^{\sss{(\ep_{\sss P}}\sss{n)}}d^{\sss{n}}u\;
    d^{\sss{n}}\tau\funcleft{K^*}{P}{(u,\tau)}{G}}\nn\\
    &&{}-({F}\leftrightarrow
    {G})(-1)^{\sss{(\ep({F})+1)(\ep({G})+1)}},
\eea
where $F$ and $G$ are two functionals of the superfields. Decomposing
 the superfields into ordinary fields by expanding in the odd
 coordinates $\tau^{\sss a}$ this reduces to the standard expression for the BV-bracket \cite{Batalin:2001ge}.

It follows that this functional bracket has the standard properties, such as the graded symmetry property
\be
  ({F},{G})=-(-1)^{\sss{(\ep({F})+1)(\ep({G})+1)}}({G},{F}),
\ee
the graded Leibniz rule
\bea
({F},{G}{H})=({F},{G}){H}+{G}({F},{H})(-1)^{\sss{\ep({G})\ep({H})}},\\
({F}{G},{H})=({F},{H}){G}+{F}({G},{H})(-1)^{\sss{\ep({F})\ep({G})}},
\eea and the graded Jacobi-identity
\be
(({F},{G}),{H})(-1)^{\sss{(\ep({F})+1)(\ep({H})+1)}}+\mbox{cycl}({F},{G},{H})=0.
\ee

Proceeding in the same way as in \cite{Batalin:2001ge,
  Batalin:2001su}, we now evaluate the left-hand side of
  (\ref{classicalmaster}) to find a simple local expression for the master equation
(\ref{classicalmaster}) that this generalized model $\Si$ has to satisfy. In order to derive this local
expression the following properties of the functional derivatives
are needed \bea
   \funcleft{K^*}{P_1}{(u,\tau)}K^*_{\sss
 {P_2}}(u',\tau')&=&\delta^{\sss{P_1}}_{\sss{P_2}}\delta^{\sss n}(u-u')\delta^{\sss n}(\tau-\tau'),\nonumber\\
 \ep(\funcleft{K^{\sss
 P}}{}{(u,\tau)})&=&\ep(\funcright{K}{P}{(u,\tau)})=\ep_{\sss P}+n,\nonumber\\
\ep(\funcleft{DK^{\sss
 P}}{}{(u,\tau)})&=&\ep(\funcright{DK}{P}{(u,\tau)})=\ep_{\sss P}+n+1.
\eea
A straightforward calculation of the master equation
(\ref{classicalmaster}) now gives
 \bea \label{requires}
0&=&\frac{1}{2}(\Si,\Si)=\frac{1}{2}\int_{\mathcal M}
d^{\sss{n}}ud^{\sss{n}}\tau(\mathcal{L}_{n},\mathcal{L}_{n})_{\sss n}\nn\\
&=& \int_{\mathcal M}
d^{\sss{n}}ud^{\sss{n}}\tau\Big(\frac{1}{2}(S,S)_{\sss
 n}+D\mathcal{L}_{\sss n}
-D\Big(DK^{\sss {P}}\leftder{S}{DK}{P}+DK^*_{\sss
  {P}}\leftder{S}{DK_{\sss{P}}}{*}\Big)\Big),\nonumber\\
&&{}
\eea
where we have introduced the following bracket for the local
functions $A(u,\tau)$ and $B(u,\tau)$
\bea \label{newbracket}
   (A,B)_{\sss n}\!\!\!\!\!\!&=&\!\!\!\!\!\!\Big(\!A\frac{\stackrel{\leftarrow}{\partial}}{\partial{K^{\sss{P}}}}+(-1)^{\sss{\ep_P+\ep(A)}}D\Big(\!\rightder{A}{DK}{P}\!\Big)\!\Big)
\Big(\!\frac{\stackrel{\rightarrow}{\partial}}{\partial{K^*_{\sss{P}}}}B+(-1)^{\sss{\ep_P+n}}D\Big(\!\leftder{B}{DK_{\sss{P}}}{*}\!\Big)\!\Big)\nn\\
&&\!\!\!\!\!\!-(A\leftrightarrow
\!B)(-1)^{\sss{(\ep(A)+n+1)(\ep(B)+n+1)}}. \eea
In order for the
master action $\Si$ to describe a possible consistent gauge field
  theory (i.e satisfy the master equation (\ref{requires})), we
  require the boundary conditions on the superfields to satisfy
 \be \label{boundary}
   \int_{\mathcal M} d^{\sss{n}}ud^{\sss{n}}\tau\Big(
   D\mathcal{L}_n(u,\tau)-D\Big(DK^{\sss {P}}\leftder{S}{DK}{P}+DK^*_{\sss {P}}\leftder{S}{DK_{\sss{P}}}{*}\Big)\Big)=0.
 \ee
Note that this condition is trivially satisfied for
manifolds without a boundary, or for Dirichlet boundary conditions
on the fields for manifolds with a boundary. From (\ref{requires})
it now follows that we may require the local function $S$ to be a solution
to the local master equation 
\be \label{mastereq} 
(S,S)_{\sss n}=0, 
\ee 
in terms of the bracket (\ref{newbracket}).
In principle, more general forms of boundary conditions than
  (\ref{boundary}) are possible in which case (\ref{mastereq}) has to be modified.

The bracket (\ref{newbracket}) reduces to the ordinary local bracket
if the local functions, such as $S$, does not depend on terms
involving $DK^{\sss P}, DK^*_{\sss P}$. In this case (\ref{newbracket}),
(\ref{boundary}) and (\ref{mastereq}) are the same as the ones
found in \cite{Batalin:2001su}. It is the new bracket
(\ref{newbracket}) and the introduction of super multiplier fields that makes it
possible to derive higher order gauge field theories from the
superfield algorithm.

The local expression (\ref{newbracket}) has the graded symmetry
property
 \be\label{gradesymm}
    (F,G)_{\sss{n}}=-(-1)^{\sss{(\ep(F)+n+1)(\ep(G)+n+1)}}(G,F)_{\sss{n}},
 \ee
carries ($1$-$n$) units of ghost number
 \be\label{ghostcarry}
    \gh(F,G)_{\sss{n}}=\gh F+\gh G+1-n,
 \ee
and ($n$+$1$) units of parity \be
   \ep((F,G)_{\sss{n}})=\ep(F)+\ep(G)+n+1.
\ee 
Due to the $D$-terms in (\ref{newbracket}) the Jacobi-identity
and the Leibniz rule for this local bracket are not easily evaluated, but since
 the bracket (\ref{newbracket}) is a local expression of the
 BV-bracket (\ref{funcbracket}) the Jacobi-identity and Leibniz rule
 of (\ref{newbracket}) are valid up to a total derivative, i.e.~modulo a $D$-term
 \cite{Barnich:1996is}. The derivations of the Leibniz rule and the Jacobi identity for higher order theories are considered in great detail in \cite{Barnich:1996is} for the bosonic case.

The equations of motion following from the master action (\ref{masteraction}) are
given by
\be \label{eom}
  DK^{\sss {P}}=(S,K^{\sss P})_{\sss n},\hspace{3mm} DK^*_{\sss
    {P}}=(S,K^*_{\sss P})_{\sss n}.
\ee
In line with \cite{Batalin:2001ge, Batalin:2001su}, the operator $D$ will now have the role of a BRST-charge
operator when acting on the superfield and the associated
superfield, since $D^2=0$. 
However, in order to have equations of motion of the form (\ref{eom})
for a general function $\Up(K^{\sss P},DK^{\sss P}, K^*_{\sss P}, DK^*_{\sss P})$ we
need to introduce a new operator $Q$ replacing $D$. This operator is
given by
\be \label{Qdef} 
Q\Up:=D\Up+D(DK^{\sss
P}\leftder{\Up}{DK}{P} +DK^*_{\sss
  P}\leftder{\Up}{DK_{\sss{P}}}{*}),
\ee 
with $\ep(Q)=1$ and $\gh Q=1$. Using the equations of motion
above we find 
\be \label{eomQ} 
Q\Up=(S,\Up)_{\sss n}, 
\ee 
which implies that $QS=(S,S)_{\sss n}$ and $Q^2S=(S,(S,S)_{\sss
n})_{\sss n}$. Note that expression (\ref{eomQ}) above reduces to
(\ref{eom}) if $\Up$ is just a superfield or an associated
superfield. $Q$ is a nilpotent operator using (\ref{eom}),
i.e.~$Q^2=0$, which gives $Q$ an interpretation of a BRST-charge
operator. Since $0=Q^2\Up=(S,(S,\Up)_{\sss n})_{\sss n}$ it follows
from the graded Jacobi identity that consistency requires that
$(S,S)_{\sss n}=0$. This argument supports our proposal (\ref{mastereq}).
Note that with the operator $Q$ we may write the conditions (\ref{boundary}) as \be
\label{Qboundary} \int_{\mathcal M}
d^{\sss{n}}ud^{\sss{n}}\tau\big(D(K^*_{\sss P}DK^{\sss
P})-QS\big)=0, \ee where the last term disappear on-shell. For
manifolds without boundary we can disregard any $D$-exact terms.
For reasonable choices of boundary conditions on the fields, such
as for example Dirichlet conditions, the $D$-exact terms vanishes
on the boundary. All of the analysis made in the rest of this
paper, assumes that such a choice has been made.

The expansion of the superfields $K^{\sss P}$ and
$K^*_{\sss P}$ in terms of the odd coordinates $\tau^{\sss{a}}$
lead to component fields which are either conventional BV fields
or -antifields. Since the original fields constitute the ghost
number zero components of the superfields $K^{\sss P}$ and
$K^*_{\sss P}$, one obtains the following rules
\cite{Batalin:2001su} for extracting the $n$-dimensional
classical field theory corresponding to a given master action
$\Si$ of the form (\ref{masteraction}) and (\ref{Lagrangedensity}):
 \bea\label{limit}
    d^{\sss n}ud^{\sss n}\tau & \rightarrow & 1\nn\\
    D & \rightarrow & \mbox{exterior derivative } d\nn\\
    K^{\sss P}:\gh K^{\sss P}=k\geq 0 & \rightarrow & \mbox{k-form
    field}\,\,k^{\sss P} \mbox{where},\nn\\
    & &\ep(k^{\sss P})  = \ep_{\sss P}\mbox{\it\fs +k}\nn\\
    K^*_{\sss P}:\gh K^*_{\sss P}=(n-1-k)\geq 0 & \rightarrow &\mbox{\fs\it(n-1-k)}-\mbox{form field}\,\,
    k^*_{\sss P}\,\mbox{where},\nn\\
    & & \ep(k^*_{\sss P})=\ep_{\sss P}\mbox{\it\fs +k}\nn\\
    \mbox{all other superfields} & \rightarrow & 0\nn\\
    \mbox{pointwise multiplication} & \rightarrow & \mbox{wedge
    product}.
 \eea
The variation of the superfields in $\Si$ are given by
\bea
\label{gaugetransf1}
   \de_{\Si}K^{\sss P}&\!:=\!&(\Si,K^{\sss P})=(-1)^{\sss n}(DK^{\sss
     {P_1}}-(S,K^{\sss P})_{\sss n}),\\
\label{gaugetransf2}
   \de_{\Si}K^*_{\sss P}&\!:=\!&(\Si,K^*_{\sss P})=(-1)^{\sss n}(DK^*_{\sss {P_1}}-(S,K^*_{\sss P})_{\sss n}),
\eea
from which one can determine the gauge transformations of a
classical model by applying the rules (\ref{limit}) in the same
way as was considered in \cite{Edgren:2002fi}.

%
%
\section{Generalized superfield algorithm}
We begin this section by considering some of the limitations
of the superfield algorithm in \cite{Batalin:2001su} when it comes to
describing higher order gauge field theories. Then we show how
the introduction of non-dynamical super multiplier fields
$\La^{\sss P}$ and $\La^*_{\sss P}$ can resolve these limitations.

Consider the reduction rules (\ref{limit}) from which the original
model of a given master action may be derived. Now, in general a master
action in which all the fields are dynamical, does not provide for the
possibility of an explicit relation
between the fields and their associated fields. However, a master
action with only dynamical fields might provide
for an implicit relation between the fields and the associated
fields. This is for instance the case for theories like topological Yang-Mills
(TYM), in which the associated field can be eliminated by
means of its own equation of motion. This implies that the
original model, i.e.~classical TYM only contain one field, namely
the gauge connection $1$-form, exactly as it should. It is easy to
see that such an implicit relation between dynamical fields and
associated fields is only possible if there exists a quadratic term
of either the field or the associated field in the master action.
Only then may the equation of motion for that field give a
sensible relation to its corresponding associated field. However,
a quadratic term is only possible in even dimensions, which rules
out such implicit relations for all models in odd dimensions.
Moreover, for many models in even dimensions such 'searched for'
relations are not possible either. The reason for this is that the
ghost number prescription of the master action requires the
ghost number of the field in the quadratic term to be equal the
dimension of the model divided by two, which is not the case for
most fields. Important odd dimensional models which cannot be
formulated without multiplier fields, includes for example $d=3$
Chern-Simons models and higher dimensional analogues - all of
which clearly contain only one field: the gauge connection
$1$-form.

Introducing non-dynamical fields into the model can be used to impose
(via their equations of motions) relations (constraints) between fields and the associated
fields in the master action. Relations which are carried over to
the original model, via the rules (\ref{limit}). Thus, a typical
multiplier term in the master action looks like
 \be
    \La^{\sss P}\phi_{\sss P}(K,K^*,DK,DK^*),
 \ee
and imposes the relation $\phi_{\sss P}=0$ between $K$ and $K^*$
through the equations of motion. Consider
now the equations of motion (\ref{eom}) following from a given
local master action $S$,
 \be
    DK=(S,K)_{\sss n}.
 \ee
Above, the indices are suppressed and $K$ denotes either a
superfield or an associated superfield. Since $S$ is linear in
the multiplier fields, two things can happen: 1) $S$ does not
contain the associated field to a given multiplier field, say
$\La$, in which case we have $(S,\La)=0$. 2) $S$ contains the
associated field to a given field, say $\La$, in which case we
have $(S,\La)=\phi$ for some constraint $\phi$. On the constraint
surface, however, we have also in this case $(S,\La)=0$. Due to the
general form of the equations of motion (\ref{eom}) we therefore
define the super multiplier fields to be those satisfying 
the following properties,
 \bea
    D\La^{\sss P} &=& 0,\\
    D\La^*_{\sss P} &=& 0.
 \eea
For higher order theories in general, the local master equation
(\ref{mastereq}) will include terms
containing the de Rham differential $D$ on some fields. If these
terms are $D$-exact they may be ignored - since they may be
absorbed by redefinitions of the local bracket $(\,\,,\,)_{\sss
n}$ (this might affect the boundary conditions). If not, we must use
the equations of motion (\ref{eom}) of the
corresponding fields in order to obtain terms which are polynomial
in the fields only. Depending on the theories considered, we
might be forced to use this procedure repeatedly in order to bring
all the terms in the local master equation on polynomial form in
the fields. Only then may we identify the equations that our
coefficients in the master action must satisfy. It is in this
sense, that the solutions found to the local master equation are
on-shell solutions. The relations imposed by the constraints
should also be imposed at this level. This is so since the
multiplier fields impose the constraints by means of their
equations of motion and we are interested in the equations that
the master equation imply for the independent dynamical fields.

The steps above may be summarized:
\begin{itemize}
    \item[1)] Calculate the local master equation $(S,S)_{\sss
    n}=0$.
    \item[2)] Impose the constraints by using the equations of
    motion for the Lagrange super multiplier fields.
    \item[3)] $D$-exact terms in the master equation may be
    set to zero since they do not contribute to the solution.
    \item[4)] Eliminate all kinetic terms that are left by using
    the appropriate equations of motion for the corresponding
    fields (with kinetic terms is here meant terms containing $D$
    on some fields).
 \end{itemize}

\noindent The main motivation for this algorithm is the
fact that it seems to work on all models tested so far. In some
cases, for example topological Yang-Mills in $n=4$ for which we
have alternative formulations, the algorithm above generates
identical solutions to the master equation as those obtained by
considering only dynamical fields.

%
%

\section{Topological Yang-Mills theory}

Topological Yang-Mills theory in four dimensions, originally
considered in \cite{Witten:1988to, Baulieu:1988to}, is defined by
the action
 \be S_{\sss TYM}=\int tr[F{{\wedge}}F]=
    \int tr[dA^{\sss 2}+2A^{\sss 2}dA+A^{\sss 4}].
 \ee
The field strength $F$ is related to the gauge connection $1$-form
$A$ as $F=dA+A^{\sss 2}$. Locally, $A=g_{\sss a}A^{\sss a}=g_{\sss
a}A^{\sss a}_{\mu}dx^{\mu}$, where $g_{\sss a}$ satisfies the Lie-algebra $[g_{\sss
  a},g_{\sss b}]={f_{\sss ab}}^{\sss c}g_{\sss c}$. The expression
above can now be written in terms of a symmetric group metric
$g_{\sss ab}:=tr[g_{\sss a}g_{\sss b}]$ and a totally
antisymmetric structure coefficient $f_{\sss abc}={f_{\sss ab}}^{\sss d}g_{\sss
cd}$ , \be \label{tymnotrace} {S}_{\sss TYM}=\int
g_{\sss ab}dA^{\sss a}dA^{\sss b}+{f_{\sss
    ab}}^{\sss d}g_{\sss dc}A^{\sss a}A^{\sss b}dA^{\sss c}.
\ee

This theory can be generated from a superfield formulation by
introducing a set of superfields with the following
characteristics,
\begin{center}
\begin{tabular}{|c|c|c|}
\hline
Superfield & Ghost number & Grassmann parity \\
\hline
$T^*_{\sss E}$    & $2$ & $0$ \\
$T^{\sss D}$      & $1$ & $1$ \\
$\La^*_{\sss D}$  & $0$ & $1$ \\
$\La^{\sss D}$    & $2$ & $0$ \\
\hline
\end{tabular}
\end{center}
The Grassmann parities of the superfields are chosen such that
they are even after reduction to the original theory, according to
the rules in (\ref{limit}). The master action we consider is
given by
 \be
    \label{masterym}
    \Si_{\sss TYM}=\int_{\mathcal M}{d^{\sss 4}ud^{\sss{4}}\tau
    (-T^*_{\sss{E}}DT^{\sss E}-S_{\sss 4}}),
 \ee
where the local master action is
 \be \label{localym} S_{\sss 4}=T^*_{\sss{E_1}}{\omega^{\sss{E_1}}}_{\sss{E_2E_3}}
    T^{\sss{E_2}}T^{\sss{E_3}}+\om_{\sss
    E_1E_2E_3E_4}T^{\sss{E_1}}T^{\sss{E_2}}T^{\sss{E_3}}T^{\sss{E_4}}+
    \La^{\sss{E_1}}(T^*_{\sss{E_1}}-g_{\sss{E_1 E_2}}DT^{\sss{E_2}}).
 \ee
Above, $(u^{\sss a}, \tau^{\sss a})$ denotes a set of Grassmann even
and odd local coordinates on a supermanifold ${\mathcal M}$. Since $S_4$ is even
it follows from the Grassmann parities of the superfields that the
coefficients ${\omega^{\sss{E_1}}}_{\sss{E_2 E_3}}$ and $
g_{\sss{E_1 E_2}}$ are even. For simplicity, we choose these
coefficients to be independent of $(u,\tau)$. As will be seen,
this restriction is natural for the limit
(\ref{limit}) of (\ref{masterym}) to be a topological Yang-Mills
theory.

The constraint relating the associated superfield $T^*_{\sss{E}}$
with the superfields $T^{\sss{E}}$ is given by \be \phi_{\sss
E_1}=T^*_{\sss{E_1}}-g_{\sss{E_1 E_2}}DT^{\sss{E_2}}, \ee and is
first-class with respect to $(\,\,,\,)_{\sss 4}$, since
 \be (\phi_{\sss E_1},\phi_{\sss
    E_2})_{\sss 4}=2D(g_{\sss
    E_1E_2})=0.
 \ee
The constraint is needed in order to reproduce the topological
Yang-Mills action (\ref{tymnotrace}) from the master action
(\ref{masterym}), after taking the limit (\ref{limit}).

The next step is to solve the local master equation
(\ref{mastereq}) for this model. This is done at the level of the
equations of motions. In other words, we have to implement
(\ref{eom}) to obtain a solution of the local master
equation.

The procedure goes as follows; first the local master equation is
calculated, thereafter the constraint $T^*_{\sss{E_1}}=g_{\sss{E_1
    E_2}}DT^{\sss{E_2}}$ is imposed and finally the $DT^{\sss
  E}$-terms are replaced by their equations of motion $DT^{\sss
  E}=-{\om^{\sss E}}_{\sss E_1E_2}T^{\sss E_1}T^{\sss E_2}-\La^{\sss E}$.
By performing the procedure, the master equation imposes the
following set of equations:
 \bea
     0\!\!&=&\!\!(g_{\sss E_5E_6}{\om^{\sss
        E_6}}_{\sss{E_1E_2}}{\om^{\sss{E_5}}}_{\sss{E_3E_4}}+g_{\sss
      E_6E_4}{\om^{\sss
        E_6}}_{\sss{E_1E_5}}{\om^{\sss{E_5}}}_{\sss{E_2E_3}}\nn\\&-&2\om_{\sss E_1E_2E_3E_4})T^{\sss E_1}T^{\sss E_2}T^{\sss E_3}\La^{\sss E_4},\nn \\
    0\!\!&=&\!\!g_{\sss EE_1}{\om^{\sss E}}_{\sss E_3E_2}\La^{\sss
    E_1}\La^{\sss E_2}T^{\sss E_3},\nn\\
    0\!\!&=&\!\!(2{\om}_{\sss{E_1 E_2 E_3E}}{\om^{\sss E}}_{\sss E_4E_5}-g_{\sss E_6E_7}{\om^{\sss E_7}}_{\sss
    E_1E_2}{\om^{\sss E_6}}_{\sss E_3E}{\om^{\sss E}}_{\sss
    E_4E_5})T^{\sss E_1}T^{\sss E_2}T^{\sss E_3}T^{\sss E_4}T^{\sss E_5}.\nn
 \eea
We observe that if we choose the coefficient $g_{\sss
E_1E_2}$ to be symmetric and invertible (with inverse $g^{\sss
E_1E_2}$) we may write the equations above as
\bea
\label{ymjacobi}
\om_{\sss{(E_1E_2 E}}{\om^{\sss{E}}}_{\sss{E_3)
    E_4}}&=&0, \\
\label{ymsymprop} \om_{\sss E_1E_2E_3}&=&-\om_{\sss
E_3E_2E_1},\\\label{consistym} {\om}_{\sss{E_1 E_2
E_3E_4}}&=&\frac{1}{2}\om_{\sss E_1E_2E}{\om^{\sss E}}_{\sss
E_3E_4}. \eea The braces indicates a graded symmetrization over
the enclosed indices. Equation (\ref{ymjacobi}) then imposes the
Jacobi-identity on the coefficients ${\om^{\sss E_1}}_{\sss
E_2E3}$, such that these may be interpreted as the structure
coefficients of some Lie-algebra. $g_{\sss E_1E_2}$ is in this
case naturally identified as the group metric of that Lie-algebra.
Since the fields $T^{\sss E}$ are Grassmann odd, we may assume
that ${\omega^{\sss{E_1}}}_{\sss{E_2
E_3}}=-{\omega^{\sss{E_1}}}_{\sss{E_3 E_2}}$ which together with
equation (\ref{ymsymprop}) implies the total antisymmetry of the
coefficients $\om_{\sss E_1E_2E_3}=g_{\sss E_1E_4}{\om^{\sss
E_4}}_{\sss E_2E_3}$. Thus, the master equation implies that
${\om^{\sss E_1}}_{\sss E_2E_3}$ are structure coefficients of a
semi-simple Lie-algebra in this case. This is to be expected from
consistency requirements, since we have assumed $g_{\sss E_1E_2}$
to be an invertible group metric. Equation (\ref{consistym}) gives
the solution for $\om_{\sss E_1E_2E_3E_4}$. Due to the
symmetrization over the $T^{\sss E}$-fields in the master action,
this term drops out as a consequence of the Jacobi-identity. This
is also the case on the classical level for $\int tr[F\we F]$,
since the cyclicity of the trace and antisymmetry of the wedge
product forces the $A^{\sss 4}$-term to vanish.

The solved for master action can now be written,
 \be \Si_{\sss TYM}=-\int_{\mathcal M} d^{\sss4}ud^{\sss
    4}\tau(T^*_{\sss{E}}DT^{\sss{E}}+T^*_{\sss{E_1}}{\omega^{\sss{E_1}}}_{\sss{E_2E_3}}T^{\sss{E_2}}T^{\sss{E_3}}+\La^{\sss E_1}(T^*_{\sss E_1}-g_{\sss E_1E_2}DT^{\sss E_2}).
 \ee

Taking the limit (\ref{limit}) reduces $T^*_{\sss E}$ to an even
two form $t^*_{\sss E}$, $T^{\sss E}$ to an even one form $t^{\sss
E}$ and $\La^{\sss E}$ to an even two form $\la^{\sss E}$. By
implementing the constraint
 \be t^*_{\sss E_1}=g_{\sss E_1
    E_2}dt^{\sss E_2},
 \ee
the following classical action is obtained
 \be \label{tymclass}
    \Si_{\sss Cl.}=-\int g_{\sss E_1E_2}dt^{\sss
    E_1}{{\wedge}}dt^{\sss E_2}+g_{\sss E_1E}{\om^{\sss E}}_{\sss
    E_2 E_3}dt^{\sss E_1}{{\wedge}}t^{\sss E_2}{{\wedge}}t^{\sss E_3},
\ee
where the coefficients $\omega_{\sss E_1 E_2 E_3}$ belongs to
a semi-simple Lie-algebra. This is exactly the topological
Yang-Mills action (\ref{tymnotrace}) if we identify $t^{\sss E}\to
A^{a}$, $\om_{\sss E_1E_2E_3}\to f_{abc}$ and $g_{\sss E_1E_2} \to
-g_{ab}$. Hence, a topological Yang-Mills theory has been
generated using a generalized superfield algorithm.

As previously discussed, the gauge transformations of the original
model can be calculated using the superfield formalism. The
$\Sigma$-variation of the superfield $T^{\sss E}$ is
 \be \delta_{\sss{\Sigma}}T^{\sss
    E}=DT^{\sss E}-({\om^{\sss E}}_{\sss{E_1
    E_2}}T^{\sss{E_1}}T^{\sss{E_2}}+\Lambda^{\sss{E}}).
\ee
Applying the reduction rules (\ref{limit}), the gauge
transformations can be derived \bea \label{topshift} \delta
t^{\sss E}&=&d\tilde t^{\sss E}-2{\om^{\sss E}}_{\sss E_1
  E_2}\tilde t^{\sss E_1}t^{\sss E_2}-\tilde{\lambda}^{\sss E}.
\eea The tilde quantities $\tilde t^{\sss E}, \tilde \la^{\sss E}$
above are $0,1$ -form gauge parameters respectively. The first two
terms in (\ref{topshift}) correspond to the covariant
  Yang-Mills derivative and the last one to the topological shift symmetry.

The topological Yang-Mills theory can also be written in a
superfield formulation using auxiliary\footnote{We define
auxiliary fields as dynamical fields that can be eliminated by
means of their own equations of motion. In our formulation the
super multiplier fields are not dynamical.} fields instead of
multiplier fields. In this case the local master action $S$ is in
comparison with (\ref{localym}) written as
\be
S'=T^*_{\sss{E_1}}T^*_{\sss{E_2}}\omega^{\sss E_1
E_2}+T^*_{\sss{E_1}}{\omega^{\sss{E_1}}}_{\sss{E_2
E_3}}T^{\sss{E_2}}T^{\sss{E_3}}, \ee where the superfield $T^{\sss
E}$ and the auxiliary associated superfield $T^*_{\sss E}$ have
the same properties as above and the coefficients are both even.

\section{$n=5$ Chern-Simons theory}
The classical action for five dimensional (pure) Chern-Simons
theory \cite{Schwarz:1978pa, Witten:1989qu} is defined by the Lagrangian density
 \be\label{d5CSdensity}
    \mathcal{L}_{\sss CS5}=tr[{dA^2A+\frac{3}{2}dAA^3+\frac{3}{5}A^5}],
 \ee
where $A=g_{\sss a}A^{\sss a}$ denotes the usual Lie-algebra
valued 1-form (gauge connection). Locally, the gauge connection is
given by $A=g_{\sss a}A^{\sss a}_{\,{\mu}}dx^{\mu}$ where the
generators $g_{\sss a}$ satisfies a Lie-algebra $[g_{\sss a},
g_{\sss b}]={f_{\sss ab}}^{\sss c}g_{\sss c}$. If we write the
density (\ref{d5CSdensity}) explicitly in terms of the structure
constants of the Lie-algebra, we get
 \be\label{d5liepart}
    \mathcal{L}_{\sss CS5}=g_{\sss abc}dA^{\sss a}dA^{\sss b}A^{\sss
    c}+\frac{3}{4}{f_{\sss cd}}^{\sss e}g_{\sss abe}dA^{\sss a}A^{\sss b}A^{\sss c}A^{\sss d}
    +\frac{3}{20}{f_{\sss ab}}^{\sss f}{f_{\sss cd}}^{\sss
    h}g_{\sss fhe}A^{\sss a}A^{\sss b}A^{\sss c}A^{\sss d}A^{\sss e}.
 \ee
Above, the rank $3$-tensor $g_{\sss abc}$ is given by $g_{\sss abc
}:=tr[g_{\sss a}g_{\sss b}g_{\sss c}]$ and it is obviously
invariant under the adjoint action of the gauge group. In general,
for Chern-Simons theories in ($2n+1$) dimensions, $g_{\sss
a_1...a_{n+1}}$ is assumed to be a rank $n+1$ symmetric tensor.
The reason for this assumption stems from the fact that,
 \be
    d{\mathcal L}^{\sss 2n+1}_{\sss {CS}}=g_{\sss a_1...a_{n+1}}F^{\sss a_1}\we...\we
    F^{\sss a_{n+1}}
 \ee
where the 2-form Yang-Mills field strength $F^{\sss a}$, projects
out the symmetric part of $g_{\sss a_1...a_{n+1}}$. However, the
symmetry of $g_{\sss a_1...a_{n+1}}$ does not follow from
its definition, as for example can be seen from the definition of
$g_{\sss abc}$ above. In \cite{Banados:1996dy,Banados:1995lo} it
is investigated how the algebraic properties of $g_{\sss
a_1...a_{n+1}}$ determines the exact nature of the constraints in
the action (\ref{d5CSdensity}) and the higher dimensional versions
thereof. One of the conclusions of their investigation is that
higher dimensional ($n\geq 5$) Chern-Simons theories generically
do possess local degrees of freedom and are therefore not
topological gauge theories. Below, we will see how the tensor
$g_{\sss abc}$ naturally enters in the superfield formulation and
what kind of algebraic constraints that are imposed on it as a consequence of
the master equation.

Now, a superfield formulation of the five dimensional Chern-Simons
theory may be obtained by introducing superfields $S^{\sss D}$,
$S^*_{\sss D}$, $\La^{\sss D}$ and $\La^*_{\sss D}$ with the
following characteristics,\newline\newline
\begin{center}
\begin{tabular}{|c|c|c|}\hline
  Superfield & Ghost number & $\mathbb{Z}_{2}-grading$ \\\hline
  $S^{\sss D}$ & 1 & 1 \\
  $S^*_{\sss D}$ & 3 & 1\\
  $\La^{\sss D}$ & 2 & 0\\
  $\La^*_{\sss D}$ & 2 & 0\\\hline
\end{tabular}
\end{center}
\vspace{5mm}
 The specific choice of the parities of the superfields above,
make sure that the fields in the original model are Grassmann even
after the limit (\ref{limit}) is taken.

The master action we consider is then given by
 \be\label{mastercs}
    \Si_{\sss CS5}=\int_{\mathcal M} d^{\sss 5}ud^{\sss
 5}\tau\mathcal{L}_5=\int_{\mathcal M} d^{\sss
    5}ud^{\sss 5}\tau(S^*_{\sss{D}}DS^{\sss{D}}-S_{5}),
 \ee
where the local master action has the form
 \bea\label{cssuperfieldaction}
    -S_5\!\!&=&\!\!\frac{3}{5}\om_{\sss{D_1D_2D_3D_4D_5}}S^{\sss{D_1}}S^{\sss{D_2}}S^{\sss{D_3}}S^{\sss{D_4}}S^{\sss{D_5}}
    +\frac{3}{2}S^*_{\sss{D_1}}{\om^{\sss{D_1}}}_{\sss{D_2D_3}}S^{\sss{D_2}}S^{\sss{D_3}}\nn\\
    &+&\La^{\sss{D_1}}(S^*_{\sss{D_1}}-g_{\sss{D_1 D_2 D_3}}DS^{\sss{D_2}}S^{\sss{D_3}}).
 \eea
It follows from (\ref{parity}) and (\ref{ghostnumber}) that the
coefficients $\om_{\sss{D_1D_2D_3D_4D_5}}$,
${\om^{\sss{D_1}}}_{\sss{D_2D_3}}$, $g_{\sss{D_1 D_2 D_3}}$ are
all even and have ghost number zero. The coefficients are chosen
to be independent of the coordinates $(u,\tau)$ and it is obvious
from (\ref{cssuperfieldaction}) which symmetries they possess with
respect to their indices. The constraint that relates the field
$S^{\sss D}$ with the associated field $S^*_{\sss D}$ is given by
 $\phi_{\sss D_1}=0$, where
 \be
    \phi_{\sss D_1}=(S^*_{\sss{D_1}}-g_{\sss{D_1 D_2 D_3}}DS^{\sss{D_2}}S^{\sss{D_3}}).
 \ee
This choice of the constraint implies that a
possible limit (\ref{limit}) of the master action (\ref{mastercs})
is the classical five dimensional Chern-Simons action
(\ref{d5CSdensity}). This is seen by solving the constraint in
(\ref{cssuperfieldaction}) and observing that we then obtain the
same polynomiality in the fields as in the classical action. It
should be pointed out, however, that there in general exist other
possible choices of the constraint. We observe that since
 \be
    (\phi_{\sss D},\phi_{\sss D'})_{\sss 5}=-2D(g_{\sss DD_1D'}S^{\sss
    D_1}+g{\sss D{D'}D_1}S^{\sss D_1})
 \ee
is $D$-exact, $\phi_{\sss D_1}$ is first-class with respect to
$(\,\,,\,)_{\sss 5}$.

Let us now consider the (local) master equation for the action
(\ref{cssuperfieldaction}).
Following the procedure of the generalized superfield algorithm we
 calculate $(S,S)_{\sss 5}=0$ and then use that
 \bea
    S^*_{\sss{D_1}} &=& g_{\sss{D_1 D_2
    D_3}}DS^{\sss{D_2}}S^{\sss{D_3}},\\
    DS^{\sss D} &=& -\frac{3}{2}{\om^{\sss D}}_{\sss
    D_1D_2}S^{\sss D_1}S^{\sss D_2}-\La^{\sss D}.
 \eea
By collecting terms with identical field structure in the master
equation we obtain three equations,
 \bea\label{masterraw}
    0&=& \om_{\sss D_1D_2D_3D_4D}\,{\om^{\sss D}}_{\sss
    D_5D_6}S^{\sss D_1}S^{\sss D_2}S^{\sss D_3}S^{\sss D_4}S^{\sss D_5}S^{\sss
    D_6}\nn\\
    &-& \frac{3}{2}g_{\sss D_7D_8D_1}{\om^{\sss D_8}}_{\sss D_2D_3}{\om^{\sss D_7}}_{\sss D_4D_9}
    {\om^{\sss D_9}}_{\sss D_5D_6}S^{\sss D_1}S^{\sss D_2}S^{\sss D_3}S^{\sss D_4}S^{\sss D_5}S^{\sss
    D_6}\nn,\\
    0 &=& \frac{1}{2}(g_{\sss D_1D_2D}{\om^{\sss D}}_{\sss
    D_3D_4}+g_{\sss D_1DD_2}{\om^{\sss D}}_{\sss
    D_3D_4})\La^{\sss{D_1}}\La^{\sss{D_2}}S^{\sss D_3}S^{\sss D_4}\nn\\
    &-&g_{\sss D_5D_1D_3}{\om^{\sss D_5}}_{\sss D_4D_2}\La^{\sss{D_1}}\La^{\sss{D_2}}S^{\sss D_3}S^{\sss
    D_4}\nn,\\
    0 &=&\om_{\sss D_1D_2D_3D_4D}\La^{\sss D}S^{\sss D_1}S^{\sss D_2}S^{\sss D_3}S^{\sss
    D_4}\nn\\
    &-& \frac{3}{2}g_{\sss D_5DD_1}{\om^{\sss D_5}}_{\sss D_2D_6}{\om^{\sss D_6}}_{\sss D_3D_4}
    \La^{\sss D}S^{\sss D_1}S^{\sss D_2}S^{\sss D_3}S^{\sss D_4}\nn\\
    &-& \frac{3}{2}g_{\sss D_5D_6D_1}{\om^{\sss D_6}}_{\sss D_2D_3}{\om^{\sss D_5}}_{\sss D_4D}
    \La^{\sss D}S^{\sss D_1}S^{\sss D_2}S^{\sss D_3}S^{\sss D_4}\nn\\
    &+& \frac{3}{2}g_{\sss DD_5D_6}{\om^{\sss D_6}}_{\sss D_1D_2}{\om^{\sss D_5}}_{\sss D_3D_4}
    \La^{\sss D}S^{\sss D_1}S^{\sss D_2}S^{\sss D_3}S^{\sss D_4}.
 \eea
Note that the polynomials in the fields above act as projection
operators, projecting out terms with certain symmetries. For
example, an equation like
 \bdm
    0=\ga_{\sss D_1D_2D_3D_4D_5D_6}
    S^{\sss D_1}S^{\sss D_2}S^{\sss D_3}S^{\sss D_4}S^{\sss
    D_5}S^{\sss D_6}
 \edm
only contains information about the totally
antisymmetric part of the coefficient $\ga_{\sss
 D_1D_2D_3D_4D_5D_6}$. We now look for the existence of solutions
 to the equations (\ref{masterraw}) in which the coefficients ${\om^{\sss
D_1}}_{\sss D_2D_3}$ belong to a semi-simple Lie-algebra. Since we
know what classical limit we would like to have from the master
action, it is clear that these are the interesting solutions from
a Chern-Simons point of view. This means in particular that the
set of ${\om^{\sss D_1}}_{\sss D_2D_3}$ will satisfy the
Jacobi-identity. Thus, we look for solutions that satisfies,
 \be
    {\om^{\sss D_5}}_{\sss (D_1|D_4}{\om^{\sss D_4}}_{\sss
    |D_2D_3)}=0.
 \ee
In the equation above the braces around the indices stand for
graded symmetrization. The Jacobi-identity implies that
the system (\ref{masterraw}) reduces to,
 \bea\label{red1}
    0&=& \om_{\sss D_1D_2D_3D_4D}\,{\om^{\sss D}}_{\sss
    D_5D_6}S^{\sss D_1}S^{\sss D_2}S^{\sss D_3}S^{\sss D_4}S^{\sss D_5}S^{\sss
    D_6},\\\label{red2}
    0 &=& \frac{1}{2}(g_{\sss D_1D_2D}{\om^{\sss D}}_{\sss
    D_3D_4}+g_{\sss D_1DD_2}{\om^{\sss D}}_{\sss
    D_3D_4})\La^{\sss D_1}\La^{\sss D_2}S^{\sss D_3}S^{\sss D_4}\nn\\
    &-&g_{\sss D_5D_1D_3}{\om^{\sss D_5}}_{\sss D_4D_2}\La^{\sss
 D_1}\La^{\sss D_2}S^{\sss D_3}S^{\sss
    D_4},\\\label{red3}
    0 &=&\om_{\sss D_1D_2D_3D_4D}\La^{\sss D}S^{\sss D_1}S^{\sss D_2}S^{\sss D_3}S^{\sss
    D_4}\nn\\
    &-& \frac{3}{2}g_{\sss D_5D_6D_1}{\om^{\sss D_6}}_{\sss D_2D_3}{\om^{\sss D_5}}_{\sss D_4D}
    \La^{\sss D}S^{\sss D_1}S^{\sss D_2}S^{\sss D_3}S^{\sss D_4}\nn\\
    &+& \frac{3}{2}g_{\sss DD_5D_6}{\om^{\sss D_6}}_{\sss D_1D_2}{\om^{\sss D_5}}_{\sss D_3D_4}
    \La^{\sss D}S^{\sss D_1}S^{\sss D_2}S^{\sss D_3}S^{\sss D_4}.
 \eea
Note that the second equation above gives an algebraic relation
between the coefficients in the constraints and the structure coefficients of
the Lie-algebra. Thus, the master equation enforces a relation
between the two a priori unrelated sets of objects $g_{\sss
D_1D_2D_3}$ and ${\om^{\sss D_1}}_{\sss D_2D_3}$.
The third equation gives a solution for the coefficient $\om_{\sss
D_1D_2D_3D_4D_5}$ in terms of the Lie-algebra coefficients and the
coefficients in the constraints. The first equation can then be
interpreted as a self-consistency condition of that solution.

Previously it was noted that the polynomials of the fields in the
master equation act as projectors and in general one might add
tensors to the solutions that vanishes under these projections. We
are however, looking for the simplest possible solutions and
therefore we do not consider this issue. Further, it is reasonable
to believe that these additional tensors in many cases can be made
to vanish by suitable redefinitions of the fields.

We will now study the solution of the equations
(\ref{red1})-(\ref{red3}) in more detail. Consider equation
(\ref{red2}), it implies that
 \be\label{grel}
    g_{\sss DD_1D_2}{\om^{\sss D}}_{\sss D_3D_4}=
    \frac{1}{2}(g_{\sss D_1D_4D}{\om^{\sss D}}_{\sss D_2D_3}+
    g_{\sss D_1DD_4}{\om^{\sss D}}_{\sss D_2D_3}).
 \ee
Inserted into (\ref{red3}) we get,
 \bea
    0 &=&\om_{\sss D_1D_2D_3D_4D}\La^{\sss D}S^{\sss D_1}S^{\sss D_2}S^{\sss D_3}S^{\sss
    D_4}\nn\\
    &-& \frac{3}{2}g_{\sss D_5D_6D_1}{\om^{\sss D_6}}_{\sss D_2D_3}{\om^{\sss D_5}}_{\sss D_4D}
    \La^{\sss D}S^{\sss D_1}S^{\sss D_2}S^{\sss D_3}S^{\sss D_4}\nn\\
    &-& \frac{3}{2}g_{\sss DD_6D_5}{\om^{\sss D_6}}_{\sss D_1D_2}{\om^{\sss D_5}}_{\sss D_3D_4}
    \La^{\sss D}S^{\sss D_1}S^{\sss D_2}S^{\sss D_3}S^{\sss D_4}\nn\\
    &+& 3g_{\sss D_6DD_1}{\om^{\sss D_6}}_{\sss D_2D_5}{\om^{\sss D_5}}_{\sss D_3D_4}
     \La^{\sss D}S^{\sss D_1}S^{\sss D_2}S^{\sss D_3}S^{\sss
     D_4}\nn.
 \eea
The last term in the system above vanishes as a consequence of the
Jacobi-identity, since the subscript indices $\scriptstyle D_2$,
$\scriptstyle D_3$ and $\scriptstyle D_4$ are symmetrized over.
This gives the solution for $\om_{\sss D_1D_2D_3D_4D}$,
 \be\label{dasolution}
    \om_{\sss D_1D_2D_3D_4D}=\frac{3}{2}\left(g_{\sss D_5D_6D_1}{\om^{\sss
    D_6}}_{\sss D_2D_3}{\om^{\sss D_5}}_{\sss D_4D}
    +g_{\sss DD_6D_5}{\om^{\sss
    D_6}}_{\sss D_1D_2}{\om^{\sss D_5}}_{\sss D_3D_4}\right).
 \ee
The solution (\ref{dasolution}) is consistent with equation
(\ref{red1}) since
 \bea
\!\!\!\!\!\!\!\!\!&&\!\!\!\!\om_{\sss D_1D_2D_3D_4D}{\om^{\sss D}}_{\sss D_7D_8}S^{\sss D_1}S^{\sss D_2}S^{\sss D_3}S^{\sss
     D_4}S^{\sss D_7}S^{\sss D_8}\nn\\\!&=&\!
    \frac{3}{2}g_{\sss DD_6D_5}{\om^{\sss D_6}}_{\sss D_1D_2}{\om^{\sss D_5}}_{\sss D_3D_4}{\om^{\sss D}}_{\sss
    D_7D_8}S^{\sss D_1}S^{\sss D_2}S^{\sss D_3}S^{\sss
     D_4}S^{\sss D_7}S^{\sss D_8}\nn\\
    \!&=&\!\frac{3}{4}g_{\sss D_6D_8D}{\om^{\sss D}}_{\sss D_5D_7}{\om^{\sss D_6}}_{\sss D_1D_2}{\om^{\sss D_5}}_{\sss
    D_3D_4}S^{\sss D_1}S^{\sss D_2}S^{\sss D_3}S^{\sss
     D_4}S^{\sss D_7}S^{\sss D_8}\nn\\
\!&+&\!\frac{3}{4}g_{\sss D_6DD_8}{\om^{\sss D}}_{\sss
    D_5D_7}{\om^{\sss D_6}}_{\sss D_1D_2}{\om^{\sss D_5}}_{\sss
    D_3D_4}S^{\sss D_1}S^{\sss D_2}S^{\sss D_3}S^{\sss
     D_4}S^{\sss D_7}S^{\sss D_8}\nn\\\!&=&\!0\nn
 \eea
as a consequence of the Jacobi-identity. A superfield
formulation of five dimensional Chern-Simons theories is thus given by the
action
 \be
    \Si=\int_{\mathcal M} d^{\sss 5}ud^{\sss 5}\tau\,{\mathcal L}_{\sss 5},
 \ee
where
 \bea
    {\mathcal L}_{5} &=& S^*_{\sss D_1}DS^{\sss D_1}\nn\\
    &+&\!\!\frac{9}{10}\left(g_{\sss D_7D_6D_1}+g_{\sss D_1D_7D_6}\right){\om^{\sss
    D_6}}_{\sss D_2D_3}{\om^{\sss D_7}}_{\sss
    D_4D_5}S^{\sss{D_1}}S^{\sss{D_2}}S^{\sss{D_3}}S^{\sss{D_4}}S^{\sss{D_5}}\nn\\
    &+& \frac{3}{2}S^*_{\sss{D_1}}{\om^{\sss{D_1}}}_{\sss{D_2D_3}}S^{\sss{D_2}}S^{\sss{D_3}}\nn\\
    &+& {}\La^{\sss{D_1}}(S^*_{\sss{D_1}}-g_{\sss{D_1 D_2 D_3}}DS^{\sss{D_2}}S^{\sss{D_3}}).
 \eea
After eliminating the super multiplier field $\La$, using relation
(\ref{grel}) and performing a partial integration of the kinetic
term, we get \bea
    {\mathcal L}_{5}|_\La &=& \frac{1}{2}\left(g_{\sss D_1D_2D_3}+g_{\sss D_2D_3D_1}\right)DS^{\sss D_2}S^{\sss D_3}DS^{\sss D_1}\nn\\
    &+&\!\!\frac{9}{10}\left(g_{\sss D_7D_6D_1}+g_{\sss D_1D_7D_6}\right){\om^{\sss
    D_6}}_{\sss D_2D_3}{\om^{\sss D_7}}_{\sss
    D_4D_5}S^{\sss{D_1}}S^{\sss{D_2}}S^{\sss{D_3}}S^{\sss{D_4}}S^{\sss{D_5}}\nn\\
    &+& \frac{3}{4}\left(g_{\sss D_2D_5D_1}
    +g_{\sss D_2D_1D_5}\right){\om^{\sss{D_1}}}_{\sss{D_3D_4}}DS^{\sss D_4}S^{\sss D_5}S^{\sss{D_2}}S^{\sss{D_3}}.\nn\\
 \eea
In order to obtain Chern-Simons theory we impose symmetry between
the first two indices of the $g$ tensor,
 \be
  g_{\sss D_1D_2D_3}=g_{\sss D_2D_1D_3}.
 \ee
This choice is very natural since if we define
 \be
    \tilde{g}_{\sss D_1D_2D_3}=g_{\sss D_1D_2D_3}+g_{\sss D_1D_3D_2},
 \ee
$\tilde{g}_{\sss D_1D_2D_3}$ will be totally symmetric. We get
thus,
\bea \label{d5Chern}
    {\mathcal L}_{5}|_\La &=& \frac{1}{2}\tilde{g}_{\sss D_1D_2D_3}DS^{\sss D_2}S^{\sss D_3}DS^{\sss D_1}\nn\\
    &+&\!\!\frac{9}{10}\tilde{g}_{\sss D_7D_6D_1}{\om^{\sss
    D_6}}_{\sss D_2D_3}{\om^{\sss D_7}}_{\sss
    D_4D_5}S^{\sss{D_1}}S^{\sss{D_2}}S^{\sss{D_3}}S^{\sss{D_4}}S^{\sss{D_5}}\nn\\
    &+& \frac{3}{4}\tilde{g}_{\sss D_2D_5D_1}{\om^{\sss{D_1}}}_{\sss{D_3D_4}}DS^{\sss D_4}S^{\sss D_5}S^{\sss{D_2}}S^{\sss{D_3}}.\nn\\
 \eea
 Note that this expression has exactly the same
structure as the one in (\ref{d5liepart}). Thus, we expect this to
be the master action of the five dimensional Chern-Simons theory
by means of an appropriate redefinition of the coefficients. In the
case of an abelian Chern-Simons theory this is obtained from
(\ref{d5Chern}) by applying the reduction rules (\ref{limit}) and
identifying $S^{\sss D_1}\to A^{\sss a}, {\om^{\sss{D_1}}}_{\sss{D_2D_3}}
\to 0, \frac{1}{2}\tilde{g}_{\sss D_1D_2D_3} \to g_{\sss abc}$.

%
%
%
%
%
%
%
%
\section{$n=3$ theories}
In this section we take a look at some three dimensional models.
When trying to formulate $n=3$ Chern-Simons theories in terms of a
master action, complications occur when one uses the superfield
algorithm. The reason for this is that the most natural (simplest)
constraints that relates superfields and associated superfields
are generically second class with respect to $(\,\,,\,)_{\sss 3}$
in this case. This implies that we obtain some restrictions on the
Lagrange multipliers $\La^{\sss F}$ that enforces the constraints.
Consider the classical action of $d=3$ Chern-Simons,
 \be\label{d3CSdensity}
    \mathcal{L}_{\sss CS3}=tr[AdA+\frac{2}{3}A^{\sss 3}].
 \ee
The notation is identical to that of the preceding section. The
characteristics of the fields are,
\begin{center}
\begin{tabular}{|c|c|c|}\hline
  Superfield & Ghost number & $\mathbb{Z}_{2}-grading$ \\\hline
  $U^{\sss F}$ & 1 & 1 \\
  $U^*_{\sss F}$ & 1 & 1\\
  $\La^{\sss F}$ & 2 & 0\\
  $\La^*_{\sss F}$ & 0 & 0\\\hline
\end{tabular}
\end{center}
The specific parities above are chosen as to produce even fields
in the limit (\ref{limit}) for the following master action:
 \be\label{cssuper}
    \Si=\int_{\mathcal M} d^{\sss 3}ud^{\sss 3}\tau(U^*_{\sss{F}}DU^{\sss{F}}-U^*_{\sss F_1}{\om^{\sss F_1}}_{\sss F_2F_3}
    U^{\sss F_2}U^{\sss F_3}-\La^{\sss F_1}(U^*_{\sss F_1}-g_{\sss F_1F_2}U^{\sss F_2})).
 \ee
The constraints above are the simplest ones possible if we want
the preceding master action to reduce to an action defined by
(\ref{d3CSdensity}). The constraints $\phi^{\sss F}$ are given by
 \be
    \phi_{\sss F_1}=U^*_{\sss F_1}-g_{\sss F_1F_2}U^{\sss F_2},
 \ee
and their algebra with respect to $(\,\,,\,)$ is given by,
 \be
    (\phi^{\sss F_1},\phi^{\sss F_2})_{\sss 3}=-g^{\sss F_1F_2}-g^{\sss F_2F_1}.
 \ee
This implies that $\phi^{\sss F_1}$ is of first-class with respect
to $(\,\,,\,)_{\sss 3}$ only if $g^{\sss F_1F_2}$ is
antisymmetric. We cannot identify $g_{\sss F_1F_2}$
with some group metric, since in that case it must be symmetric.
Without making any assumptions about the symmetries of the
$g_{\sss F_1F_2}$ coefficient, we now calculate the master
equation,
 \bea\label{d3master}
    0=(S,S)_{\sss 3} &=& 2 g{\sss F_1F_5}U^{\sss F_5}{\om^{\sss F_1}}_{\sss F_2F}{\om^{\sss F}}_{\sss
    F_3F_4}U^{\sss F_2F_3F_4}\nn\\
    &+& 2 g_{\sss F_1F_5}{\om^{\sss F_1}}_{\sss F_2F}U^{\sss F_5}U^{\sss F_2}\La^{\sss
    F}\nn\\
    &-& \La^{\sss F_1}g_{\sss F_1F}{\om^{\sss F}}_{\sss
    F_3F_4}U^{\sss F_3}U^{\sss F_4}\nn\\
    &-& \La^{\sss F_1}g_{\sss F_1F}\La^{\sss F},
 \eea
which gives the following three equations,
 \bea
    0 &=& g_{\sss F_1F_2}+g_{\sss F_2F_1},\\
    0 &=& \om{\sss (F_5F_2|F}{\om^{\sss F}}_{\sss |F_3F_4)},\\
    0 &=& \om_{\sss F_5F_2F}-\om_{\sss F_2F_5F}-\om_{\sss FF_5F_2}\label{gi}.
 \eea
Note that in the first equation above, the first-class requirement
is imposed as a condition via the master equation. Thus, if the
model described by (\ref{cssuper}) is to describe a consistent
gauge theory, $g_{\sss F_1F_2}$ has to be antisymmetric. The
coefficients $\om_{\sss F_5F_2F}=g_{\sss F_5F_6}{\om^{\sss
F_6}}_{\sss F_2F}$ cannot belong to a semi-simple Lie-algebra in
this case and it is precisely that fact that makes the last
equation above non-trivial (otherwise we would have $\om_{\sss
F_1F_2F_3}=0$). The second equation can be viewed as a generalized
Jacobi-identity. The classical limit of the theory under
consideration is, however in the first-class scenario, given by
the non-dynamical model,
 \be\label{nondym}
    \int \om_{\sss F_1F_2F_3}u^{\sss F_1}\we u^{\sss F_2}\we u^{\sss F_3}.
 \ee
The reason for this is that if $g_{\sss F_1F_2}$ is
anti-symmetric, we have (for suitable boundary conditions on the
fields $u^{\sss F}$)
 \be
    \int g_{\sss F_1F_2}u^{\sss F_1}\we du^{\sss F2}=-\int g_{\sss
    F_2F_1}u^{\sss F_2}\we du^{\sss F_1}.
 \ee
The action (\ref{nondym}) is gauge invariant as a consequence of
equation (\ref{gi}) since,
 \bea
    \de(\om_{\sss F_1F_2F_3}u^{\sss F_1}\we u^{\sss F_2}\we u^{\sss F_3}\we
    )&=&
     (\om_{\sss F_1F_2F_3}-\om_{\sss F_2F_1F_3}+\om_{\sss
    F_3F_2F_1})\de u^{\sss F_1}\we u^{\sss F_2}\we u^{\sss
    F_3}.\nn
 \eea
There exists also other solutions to this model, when the
first-class requirement with anti-symmetric $g_{\sss F_1F_2}$ is
not satisfied. From (\ref{d3master}) we see that we always have
the trivial solution $\La^{\sss F}=0$, which amounts to a three
dimensional model with a Lie-algebra symmetry. Another, more
interesting possibility is to consider the master equation as
imposing constraints on the super multiplier fields $\La^{\sss F}$.
It is then easy to see directly from (\ref{d3master}) that if we
set
 \be
    \La^{\sss F} = {\om^{\sss F}}{\sss F_1F_2}U^{\sss F_1}U^{\sss F_2}
 \ee
we have a consistent solution to the master equation, provided the
coefficients ${\om^{\sss F}}{\sss F_3F_4}$ satisfies the
Jacobi-identity. In this case, the coefficient $g_{\sss F_1F_2}$
may have a group metric interpretation. The classical limit of
such a model is given by
 \be
    \int u^*_{\sss F}\we du^{\sss F} + 2\,{\om^{\sss F}}{\sss
    F_3F_4}\;u^*_{\sss F}\we u^{\sss F_3}\we u^{\sss F_4}-\om{\sss
    F_1F_2F_3}\;u^{\sss F_1}\we u^{\sss F_2}\we u^{\sss F_3}.
 \ee
\vspace{5mm}

\noindent
{\bf Acknowledgments:}
We would like to thank R. Marnelius for valuable comments.

\bibliographystyle{utphysmod2}
\bibliography{biblio1}
\end{document}